\documentstyle[12pt]{article}
\begin{document}\begin{flushright}\thispagestyle{empty}
OUT--4102--81\\
13 September 1999\\
hep-ph/9909336                                     \end{flushright}\vfill
                                                   \begin{center}{ \Large\bf
Four-loop Dyson-Schwinger-Johnson anatomy       }\vglue 10mm{\large\bf
D.~J.~Broadhurst                                   }\vglue 5 mm{\large
Physics Department, Open University,
Milton Keynes MK7 6AA, UK                          \\[3pt]{\tt
http://physics.open.ac.uk/$\;\widetilde{}$dbroadhu \\[3pt]
D.Broadhurst@open.ac.uk                            }\vfill

{\sl In memoriam\/} Kenneth Johnson 1931--1999    }\end{center}\vfill
                                                  \noindent{\bf Abstract}\quad
Dyson-Schwinger equations are used to evaluate the 4-loop anomalous dimensions
of quenched QED in terms of finite, scheme-independent, 3-loop integrals.
Three of the results serve as strong checks of terms from scheme-dependent
4-loop QCD calculations. The fourth, for the anomalous dimension of
$\overline\psi\sigma_{\mu\nu}\psi$, was previously known only to 2-loop order.
The 4-loop beta function has 24 unambiguous terms. Two of the simplest give
$\beta_4=\frac13(22-160)=-46$. The rational, $\zeta_3$ and $\zeta_5$ parts of
the other 22 miraculously sum to zero. Vertex anomalous dimensions have 40
terms, with no dramatic cancellations. Our methods come from work by the late
Kenneth Johnson, done more than 30 years ago. They are entirely free of the
subtractions and infrared rearrangements of later methods.
\vfill{}\newpage\setcounter{page}{1}
\newcommand{\df}[2]{\mbox{$\frac{#1}{#2}$}}

\subsubsection*{1.\quad Introduction}

Over 30 years ago,
Kenneth Johnson, Raymond Willey and Marshall Baker~\cite{JWB} probed
perturbative QED as deeply as was allowed by current analytical methods.
Since then, analysis has progressed from three loops~\cite{JWB,JB,JLR,DRR} to
four~\cite{nbg,4LQ,GKL,beta,Kostja,mass}. Arguably,
this technical advance may
have been at the expense of field-theoretic insight from QED.
Here we show how the Dyson-Schwinger equations~\cite{JWB,BD,BDK}
of quenched QED provide highly non-trivial tests of terms in the
4-loop anomalous dimensions of QCD.

The computational cost is so low that one
may dispense with {\sc mincer}~\cite{mincer},
and use its humble relation {\sc slicer}~\cite{BKT},
which implements the 3-loop methods of~\cite{CT} in {\sc reduce}.
This increases the independence of our tests.
No labour is expended on infrared~\cite{IRA} rearrangement.
Our scheme-independent results for the 4-loop coupling and mass
confirm those obtained by minimal subtraction~\cite{4LQ,Kostja,mass}.
We compute the 4-loop anomalous tensor dimension,
previously known to 2 loops~\cite{MM,BG} and expose the
Dyson-Schwinger-Johnson anatomy of the 4-loop beta term~\cite{4LQ,beta,BK},
$\beta_4=-46$, which is awesome to behold.

\subsubsection*{2.\quad The kernel of the matter}

The anomalous dimensions $\beta(\alpha_s)$ and $\gamma_m(\alpha_s)$
of the coupling and fermion mass have been computed to 4 loops
for an SU$(N)$ gauge theory with an arbitrary number of flavours, $N_f$.
In the case of $\beta$, which involves $O(10^4)$ diagrams, only one
group~\cite{beta} has been able to achieve a result; in the case of
$\gamma_m$, two analyses~\cite{Kostja,mass} agree.
Here we test contributions common to quenched QED, i.e.\ the
leading terms in the limit $N_f\to0$, $N\to0$, with
$\alpha=\alpha_s(N^2-1)/2N$ held constant.

The key~\cite{JWB} to QED is $K$, the
one-photon-irreducible, two-fermion-irreducible,
electron-positron kernel: the only
{\em perturbative\/} construct in the
Dyson-Schwinger equations~\cite{BD}
\begin{eqnarray}
\Sigma&=&e^2Z_1\Gamma_\mu S D^{\mu\nu}\gamma_\nu\label{DS1}\\
\Pi_{\mu\nu}&=&e^2Z_1\Gamma_\mu G\gamma_\nu\label{DS2}\\
\Gamma_\mu&=&Z_1\gamma_\mu+\Gamma_\mu GK\label{DS3}
\end{eqnarray}
The fermion propagator is $S(p)$. Its wave-function and mass
renormalization constants,
$Z_2$ and $Z_m$, absorb linear and logarithmic divergences from the
self-energy $\Sigma(p)$ in
\begin{equation}
S^{-1}(p)=Z_2(p\llap{/\kern-0.5pt}-m)-\Sigma(p)\label{S}
\label{Si}
\end{equation}
with a bare mass $m_0=m/Z_m$.
Similarly, $Z_3$ absorbs the logarithmic
divergences in $\Pi_{\mu\nu}(k)=(k^2g_{\mu\nu}-k_\mu k_\nu)\Pi(k)$,
giving a photon propagator
\begin{equation}
D_{\mu\nu}(k)=\frac{g_{\mu\nu}/k^2-k_\mu k_\nu/k^4}{Z_3-\Pi(k)}
+\xi k_\mu k_\nu/k^4
\label{D}
\end{equation}
with a gauge parameter $\xi$ that is unaffected by
the purely transverse photon self-energy.
The Ward identity, $Z_1=Z_2$, implies a bare coupling
$e_0^2=e^2/Z_3$. The system is driven by the $GK$
term in~(\ref{DS3}), with a pair of fermion propagators
abbreviated by $G$, and the $e^+e^-$ kernel by $K$. A spin sum,
a one-loop momentum integration, and a factor of $1/(2\pi)^4$,
are to be understood in each of the convolutions effected by a pair
of propagators in~(\ref{DS1}--\ref{DS3}).

We shall work to lowest order in the number of fermion flavours,
i.e.\ in the quenched approximation, where one retains the smallest
possible number of fermion loops.
Thus we retain only those kernel terms that have precisely
one electron, precisely one positron, and at least one photon,
in every intermediate state. We omit, for example, annihilation terms with
3-photon intermediate states.
The recipe for constructing $K$ is very simple~\cite{BD}: working to
order $a^{n}$, with $a\equiv(e/4\pi)^2=\alpha/4\pi$,
one draws all the divergence-free quenched
skeletons with up to $n$ photons. These are
dressed by $\Gamma_\mu$ and $S$, while
retaining the bare photon propagator,
$g_{\mu\nu}/k^2+(\xi-1)k_\mu k_\nu/k^4$. Then $GK$
is scheme-independent for $m=0$.

Let $k_n$ be the number of
$n$-photon quenched skeletons. Quenched Wick contractions generate
$s_n=(2n-1)!!=(2n)!/2^n n!$ fermion-propagator diagrams
with $n$ loops. We form the generator $s(a)\equiv1+\sum_{n>0}s_n a^n$,
as a formal power series in the coupling $a$.
Then fermion self-energy diagrams
are generated by $1-1/s(a)$. Let $g_n$ be the number of
$n$-loop quenched vertex diagrams, generated by
$g(a)\equiv1+\sum_{n>0}g_n a^n$. Dyson equation~(\ref{DS1}) requires that
$1-1/s(a)=a g(a)s(a)$, which yields
the integer sequence
\begin{equation}
1,\,6,\,50,\,518,\,6354,\,89782,\,1435330,\,25625910,\,505785122,\,
10944711398\ldots
\label{gn}
\end{equation}
for $g_n$, revealing an error in the seventh and final entry of
sequence {\tt A005416} of Neil Sloane's on-line encyclopedia,
{\tt http://www.research.att.com/$\,\widetilde{}\,$njas/sequences}.
Then Dyson equation~(\ref{DS2}) shows that photon self-energy diagrams are
generated by $s(a)-1$. Finally, Dyson equation~(\ref{DS3}) requires that
\begin{equation}
g(a)=1+g(a)\sum_{n>0} k_n\left[a g^2(a)s^2(a)\right]^n
=\frac{s(a)-1}{as^2(a)}
\label{skn}
\end{equation}
Solving for $k_n$, we obtain the novel (and rather alarming) sequence
\begin{equation}
1,\,1,\,7,\,63,\,729,\,10113,\,161935,\,2923135,\,58547761,\,
1286468225\ldots
\label{kn}
\end{equation}

Our methods involve the zero-momentum-transfer vertex
\begin{equation}
\Gamma_0^\mu(p)\equiv\Gamma^\mu(p,p)
=\frac{\partial S^{-1}(p)}{\partial p_\mu}
\label{Ward}
\end{equation}
Hence we do not need to distinguish kernel terms related by
charge conjugation.
With 3 photons, one of the $7=1+2\times3$ skeletons is self-conjugate.
Thus there are $4=1+3$ essentially different 3-photon terms.
With 4 photons, 5 of the $63=5+2\times29$ skeletons are self-conjugate.
Thus there are $34=5+29$ essentially different 4-photon terms.
To identify each term, we give the Gauss code of the chord diagram
obtained by closing the fermion lines on the left and right
of the kernel. Starting from the closure on the left,
we pass through $2n$ vertices, on a circle, labelling
them by the $n$ chords that connect pairs of vertices.
The location of the closure on the right is indicated by a semicolon.
Thus the $n=1$ term is $K_1\equiv K(1;1)$, the $n=2$ term is
$K_2\equiv K(1,2;1,2)$,
and the $n=3$ terms are
$K_{3a}\equiv K(1,2,3;1,2,3)$, $K_{3b}\equiv K(1,2,3;1,3,2)$,
$K_{3c}\equiv K(1,2;3,1,2,3)$ and $K_{3d}\equiv K(1,2;3,2,1,3)$.
The 34 terms with $n=4$ photons are listed later.
The 4-loop scalar, vector and tensor vertices involve
$1+1+4+34=40$ kernel terms. First we consider the $\beta$ function,
which involves only $1+1+4=6$ kernel terms.

\subsubsection*{3.\quad Four-loop $\beta$ function of quenched QED}

A subdivergence-free Dyson-Schwinger method for the $\beta$ function
was developed in~\cite{BDK}, following the lead of~\cite{JWB}.
We expand the vertex
$\Gamma^\mu=\Gamma^\mu_0+k^\alpha\Gamma^\mu_\alpha+O(k^2)$
to first order in the photon momentum $k$, with subscript $\alpha$
denoting differentiation w.r.t.\ $k^\alpha$.
Next, we doubly differentiate the $O(k^2)$ combination
$(\Gamma^\mu-\Gamma^\mu_0)G(1-KG)(\Gamma^\nu-\Gamma^\nu_0)$
and make much use of the identity~\cite{JWB}
$\Gamma^\nu_\beta=(1-KG)^{-1}(KG_\beta+K_\beta G)\Gamma^\nu$,
to prove that~\cite{BDK}
\begin{eqnarray}
\Pi^{\mu\nu}_{\alpha\beta}(k)
&=&e^2\Gamma^\mu_0\left[
\df{1}{2}G_{\alpha\beta}
+G_\alpha K G_\beta
+2G_\alpha K_\beta G
+\df12GK_{\alpha\beta}G
\right.\nonumber\\&&\left.{}
+(G_\alpha K+GK_\alpha)G(1-KG)^{-1}(KG_\beta+K_\beta G)
\right]\Gamma^\nu_0
+(\alpha\leftrightarrow\beta)
+O(k)\qquad{}
\label{beta}
\end{eqnarray}
all of whose terms are subdivergence-free.
Finally, we nullify the mass and external momentum and cut
each term next to a dressed zero-momentum-transfer vertex.
Thus we obtain the 4-loop beta function from 24 finite, scheme-independent,
dressed two-point functions, each with no more than 3 loops.
One is a no-loop, no-kernel term; 5 are products of kernel terms;
18 result from 6 kernel terms and their first and second derivatives.

It is instructive to compare our dressed method with
the bare method~\cite{nbg,4LQ}. The latter
entails subtracting, from {\em each\/} of the $105$ bare 4-loop diagrams,
up to 8 counterdiagrams, with fewer loops. In these,
subdivergences are shrunken to points and replaced
by scheme-dependent singular counterterms, in $d=4-2\varepsilon$ dimensions.
For a general momentum routing,
each of the 105 diagrams generates 64 terms, when doubly
differentiated. For each of the 6720 bare derivatives,
one must skillfully identify an infrared-safe cutting point and apply
the identical cut to corresponding derivatives of counterdiagrams.
Then one is allowed to nullify the mass and the external momentum.
After maximizing savings from symmetries and momentum routings,
a hard core of such infrared~\cite{IRA} rearrangements finally~\cite{4LQ}
yielded $\beta(a)\equiv\partial\log(a)/\partial\log(\mu^2)
=\sum_{n>0}\beta_n a^n=\df43a+4a^2-2a^3-46a^4+O(a^5)$.
Sergei Larin kindly showed me extensive records of how this was
carefully done by hand.

The Dyson-Schwinger-Johnson decomposition~(\ref{beta}) abolishes, utterly,
the need for infrared rearrangement. At a stroke, it distributes
6720 bare derivatives into 24 classes, none of which requires subtractions.
The cut is automatically identified for each class. It is
next to a dressed zero-momentum-transfer vertex
and is totally safe. Moreover, the analysis for the 12
classes with 3-photon skeletons can be performed in 4 dimensions.
For the other 12, dimensional regularization
-- without any subtraction --
is a convenient way of respecting the Ward identity $Z_1=Z_2$.
Setting $d=4$ at the end, each class is bound
to give a finite result identical to that
from any other regularization.

In the massless theory, the Ward identity~(\ref{Ward}) gives
\begin{equation}
p^2S(p)\Gamma^\mu(p,p)=p\llap{/\kern-0.5pt}\gamma^\mu-2p^\mu\gamma(a,\xi)
\label{sg}
\end{equation}
in terms of the anomalous field dimension
\begin{equation}
\gamma(a,\xi)\equiv\frac{\partial\log(S(p)p\llap{/\kern-0.5pt})}
{\partial\log(p^2)}=\xi a+\sum_{n>1}\gamma_n a^n=
\left\{\xi-\df32a(1-a)\right\}a+O(a^4)
\label{ga}
\end{equation}
which is gauge-dependent only at one-loop order.
A no-loop calculation gives
\begin{equation}
\beta_0(a,\xi)=\df43a\left\{1-\df32\gamma^2+\gamma^3\right\}
\label{free}
\end{equation}
for the contribution of the kernel-free term to $\beta(a)$.
In Feynman gauge, $\xi=1$,
\begin{eqnarray}
\beta_0(a,1)&=&\df43a-2a^3+\df{22}{3}a^4-\df{33}{2}a^5+O(a^6)\label{bgf}\\
\beta(a)-\beta_0(a,1)&=&4a^2+O(a^4)\label{bdiff}
\end{eqnarray}
with $K_1$ adding $4a^2$ at 2 loops, whatever~\cite{JWB} gauge is chosen.

At 3 loops,~(\ref{bdiff}) reveals an unforseen circumstance:
the kernel terms sum to {\em zero\/}~\cite{JB} in Feynman gauge.
This remains as big a mystery as it was to Ken Johnson,
when he told it to me at SLAC, in 1972.
Table~1 gives the breakdown of kernel contributions to $3\beta_3$
in Feynman gauge. They sum to zero in a rather non-trivial way. From
this big zero, it follows that the 3-loop term, $-2a^3$,
first obtained by Rosner~\cite{JLR} in Landau gauge,
is given by the no-kernel Feynman-gauge contribution in~(\ref{bgf}).
For the latter, one requires only the leading term in~(\ref{ga}),
which is also given by a no-loop calculation.
If someone could prove the nullity of Table~1, without resorting
to calculation, Rosner's heroic 1966 work would dwindle to Feynman-gauge
tree diagrams. In 33 years, no-one has succeeded.

This extraordinary state of affairs motivated me
to extend the Feynman-gauge Dyson-Schwinger analysis to 4 loops.
As in~\cite{JWB} and in Table~1, the external momentum
was shared equally between the fermions, before differentiation,
and~(\ref{beta}) was contracted with $g^{\alpha\beta}g_{\mu\nu}$.
Table~2 shows the Feynman-gauge contributions to $3\beta_4$
of the 23 kernel terms.

To my mind, the tally is as staggering as for Table~1:
the sum of the first 22 terms in Table~2 {\em vanishes}.
Hence the Feynman-gauge kernel terms give
$3\beta_4(K_2^\prime)=-5\times2^5$, which one adds to $3\beta_{4,0}=
4(1-3\gamma_2)$ from the no-kernel term. Using $\gamma_2=-\frac32$, we obtain
\begin{equation}
\beta_4=\beta_{4,0}+\beta_4(K_2^\prime)=\df13(22-160)=-46\label{46}
\label{beta4}
\end{equation}
This is the first {\sc mincer}-free confirmation of the result in~\cite{4LQ},
which corrected a mistaken claim~\cite{nbg} that $\beta_4$
contains $\zeta_k\equiv\sum_{n>0}1/n^k$, with $k=3$ and $k=5$.
We have given further evidence in favour of the rational
value~(\ref{beta4}) in~\cite{BKT,BK}.

It took the humble {\sc slicer}~\cite{BKT} a CPUday to show
that the first 22 terms of Table~2 sum to zero.
By contrast, it took only two minutes to compute the
$K_2^\prime$ term. As in the 3-loop case, an ability to predict cancellations
would have simplified the calculation, vastly. One lives in hope
that the presumably~\cite{BDK} rational 5-loop quenched QED $\beta$
function will become accessible, when one has learnt how to decode
Johnson's observation~\cite{JB} of the nullity of Table~1,
and the now revealed cancellations in Table~2.

\subsubsection*{4.\quad Four-loop anomalous vertex dimensions of quenched QED}

Konstantin Chetyrkin generously supplied me with his 4-loop
expression for $Z_2$, in the minimal subtraction scheme, which he
had used en route to the result in~\cite{Kostja} for
$\gamma_m(\alpha_s)$ with $N$ colours and $N_f$ flavours. Looking at the
terms of lowest order in $N$ and $N_f$, I noted that they imply
a value of $\gamma_4$ in~(\ref{ga}) that includes both $\zeta_3$
and $\zeta_5$, though $\zeta_3$ was absent from the 3-loop result.
Since~(\ref{sg}) implies the scheme-independence of $\gamma_4$,
one ought to be able to check Chetyrkin's result without making
subtractions.

A method is given in~\cite{JWB}. Contract $GK$ on the left and right
with $\gamma_\mu$ and divide by Tr$(\gamma_\mu^2)=16$. Set
the fermion momenta on the right to zero and multiply by $p^4$,
where $p$ and $-p$ are the momenta on the left. The result is a
dimensionless, finite, scheme-independent function of the
coupling $a$ and the gauge parameter $\xi$:
\begin{equation}
G_V(a,\xi)=\xi a+\df12(\xi^2-6\xi-3)a^2+3(\xi^3+5\xi+2)a^3+O(a^4)
\label{gv}
\end{equation}
which is easily evaluated to 2 loops, i.e.\ order $a^3$.
Now one solves $G_V=0$, perturbatively, for $\xi$. The result must be
$\xi=-\sum_{n>1}\gamma_n a^{n-1}$, since the vanishing of $G_V$ implies
the vanishing of $\gamma=\xi a+\sum_{n>1}\gamma_n a^n$,
whose gauge dependence is merely at first order.

By this wonderfully simple device, one extracts $\gamma$ at $n+1$
loops from solving $G_V=0$ at $n$ loops, using
the vector projection of the
finite scheme-independent construct $GK$, nullified
on the right, with total infrared safety.
By contrast, bare perturbation theory would require one to make
scheme-dependent subtractions and skilful infrared rearrangements
of 518 bare 4-loop vertex diagrams.
In quenched QED, such labour is quite unnecessary.

Table~3 gives the contribution of the 40 kernel terms
to $\gamma_4$, in the Johnson gauge $\xi=\frac32a(1-a)$ that
makes $G_V=\gamma_4a^4+O(a^5)$. The contribution of $K_1\equiv K(1;1)$
vanishes, since the Ward identity requires that
\begin{equation}
p^2S(p)\Gamma^\mu(p,0)=(p\llap{/\kern-0.5pt}\gamma^\mu-p^\mu)
f(a)+p^\mu
\label{sg0}
\end{equation}
The radiative corrections are only in the transverse
form factor, $f(a)$, which makes no contribution to the vector
spin-projection of $GK_1$. The remaining 39 terms in Table~3
are highly non-trivial. Their total satisfyingly agrees
with Chetyrkin's private communication.

We can generalize the argument of~\cite{JWB} to give a procedure
for calculating the anomalous dimension of a scalar or tensor vertex.
In the scalar case, we contract on the left and right
with the unit matrix and divide by
Tr$(1)=4$; in the tensor case we contract with
$\sigma_{\mu\nu}\equiv\frac{i}{2}[\gamma_\mu,\gamma_\nu]$
and divide by Tr$(\sigma_{\mu\nu}^2)=48$. The 2-loop results are
\begin{eqnarray}
G_S&=&(\xi+3)a-(\xi+3)^2a^2+\left\{3\xi^3+21\xi^2+37\xi+69-12
(\xi+3)\zeta_3\right\}a^3\label{gs}\\
G_T&=&(\xi-1)a+(\xi^2-2\xi+9)a^2
+\left\{3\xi^3-7\xi^2+\df73\xi-\df{173}{3}
+4(5\xi+11)\zeta_3\right\}a^3\label{gt}
\end{eqnarray}
In the scalar case, $\gamma_m+\gamma=O(a^4)$ for $\xi=-3(1+22a^2)$.
In the tensor case, $\gamma_\sigma+\gamma=O(a^4)$ for
$\xi=1-8a+(\frac{178}{3}-64\zeta_3)a^2$,
where $\gamma_\sigma$ is the anomalous dimension of
$\overline\psi\sigma_{\mu\nu}\psi$.
To all orders, the $K_1$ contributions to $G_S$ and $G_T$ are
$(\xi+3f^2)a$ and $(\xi-f^2)a$. Hence, at 4 loops,
we need the gauge-invariant, scheme-independent
3-loop radiative corrections in
\begin{equation}
f(a)=1-2a+9a^2+\df13\left\{-247+640\zeta_5-496\zeta_3\right\}a^3+O(a^4)
\label{fa}
\end{equation}
easily obtained by~{\sc slicer}.
Table~4 gives the contributions of the 40 kernel terms to the
4-loop term in $\gamma_m+\gamma$, obtained from $G_S$
with $\xi=-3(1+22a^2)$. Hence
\begin{equation}
\gamma_m(a)+\gamma(a,\xi)=
(\xi+3)a+66a^3+\left\{-286+640\zeta_5-736\zeta_3\right\}a^4+O(a^5)
\label{2pm}
\end{equation}
with a vanishing 2-loop term. From Table~3 we obtain
\begin{equation}
\gamma(a,\xi)=
\xi a-\df32a^2+\df32a^3+
\left\{-\df{1027}{8}+80\left[8\zeta_5-5\zeta_3\right]\right\}a^4+O(a^5)
\label{gamma4}
\end{equation}
for the anomalous vertex (and hence field) dimension. Hence we find
\begin{equation}
\gamma_m(a)=3a+\df32a^2+\df{129}{2}a^3+
\left\{-\df{1261}{8}-336\zeta_3\right\}a^4+O(a^5)
\label{mass4}
\end{equation}
in agreement with~\cite{Kostja,mass}.

The tensor calculation is the simplest. The
4-photon kernel terms are needed only in Feynman gauge,
which is $2^4=16$ times faster than other gauges.
For the anomalous dimension of
$\overline\psi\sigma_{\mu\nu}\psi$, we obtain
\begin{equation}
\gamma_\sigma(a)=-a+\df{19}{2}a^2+\left\{-\df{365}{6}
+64\zeta_3\right\}a^3+\left\{\df{10489}{24}-\df{400}{3}\left[8\zeta_5
-5\zeta_3\right]\right\}a^4+O(a^5)
\label{sigma4}
\end{equation}
with the same 4-loop combination of $\zeta_5$ with $\zeta_3$ as was found
in~(\ref{gamma4}). It appears that this anomalous dimension was
previously known only at 2 loops~\cite{MM,BG}.

\subsubsection*{5.\quad Conclusions}

The 4-loop quenched QED
results~(\ref{beta4},\ref{gamma4},\ref{mass4},\ref{sigma4})
were obtained from finite
3-loop integrals, with
no scheme dependence.
The Dyson-Schwinger-Johnson anatomies of Tables~1--4
are equally unambiguous.
No subtraction of divergences was performed.
No infrared problem was encountered upon nullification
of external momenta. Such was the skill of field theorists~\cite{JWB}
thirty years ago. Since then, we have merely become better at integration.

We confirm the quenched abelian terms
in the QCD results~\cite{beta,Kostja,mass}, which had been
obtained by minimal subtractions. Moreover this confirmation
was achieved without using~{\sc mincer}, which had
until now been the sole engine for 4-loop counterterms.
The tensor anomalous dimension~(\ref{sigma4}) extends a previous
result~\cite{MM,BG} from 2 loops to 4. It was easy to obtain,
since the Johnson gauge for the 3-loop integrals is the Feynman gauge.

Were one able to understand the awesome cancellations of Tables~1 and~2,
one might hope to compute a rational result for the 5-loop $\beta$ function
of quenched QED.

\noindent{\bf Acknowledgments:}
I am most grateful to Kostja Chetyrkin for sending me his complete
results for $Z_2$ to 4 loops, in the minimal subtraction scheme.
Conversations with Ken Johnson, Andrei Kataev, Sergei Larin,
and Eduardo de Rafael fueled my fascination for quenched QED.
Three-loop work with Bob Delbourgo and Dirk Kreimer,
and hospitality from Jos Vermaseren
in Amsterdam, encouraged me to complete the 4-loop analysis.

\newpage
\raggedright


\begin{thebibliography}{99}

\bibitem{JWB}
K.~Johnson, R.~Willey, M.~Baker,
{\sl Phys.~Rev.} {\bf 163} (1967) 1699.

\bibitem{JB}
K.~Johnson, M.~Baker,
{\sl Phys.~Rev.} {\bf D8} (1973) 1110.

\bibitem{JLR}
J.L.~Rosner,
{\sl Phys.~Rev.~Lett.} {\bf 17} (1966) 1190;
{\sl Ann.~Phys.} {\bf 44} (1967) 11.

\bibitem{DRR}
E.~de Rafael, J.L.~Rosner,
{\sl Ann.~Phys.} {\bf 82} (1974) 369.


\bibitem{nbg}
S.G.~Gorishny, A.L.~Kataev, S.A.~Larin,
{\sl Phys.~Lett.} {\bf B194} (1987) 429.

\bibitem{4LQ}
S.G.~Gorishny, A.L.~Kataev, S.A.~Larin, L.R.~Surguladze,\\
{\sl Phys.~Lett.} {\bf B256} (1991) 81.

\bibitem{GKL}
S.G.~Gorishny, A.L.~Kataev and S.A.~Larin,
{\sl Phys.~Lett.} {\bf B259} (1991) 144.

\bibitem{beta}
T.~van Ritbergen, J.A.M.~Vermaseren, S.A.~Larin,
{\sl Phys.~Lett.} {\bf B400} (1997) 379.
{\tt hep-ph/9701390}

\bibitem{Kostja}
K.G.~Chetyrkin,
{\sl Phys.~Lett.} {\bf B404} (1997) 161.
{\tt hep-ph/9703278}

\bibitem{mass}
J.A.M.~Vermaseren, S.A.~Larin, T.~van Ritbergen,
{\sl Phys.~Lett.} {\bf B405} (1997) 327.
{\tt hep-ph/9703284}

\bibitem{BD}
J.D.~Bjorken, S.D.~Drell,
{\sl Relativistic Quantum Fields}
(McGraw-Hill, 1965).

\bibitem{BDK}
D.J.~Broadhurst, R.~Delbourgo, D.~Kreimer,
{\sl Phys.~Lett.} {\bf B366} (1996) 421.
{\tt hep-ph/9509296}

\bibitem{mincer}
S.G.~Gorishny, S.A.~Larin, L.R.~Surguladze, F.V.~Tkachov,
{\sl Comput.~Phys.~Commun.} {\bf 55} (1989) 381.

\bibitem{BKT}
D.J.~Broadhurst, A.L.~Kataev, O.V.~Tarasov,
{\sl Phys.~Lett.} {\bf B298} (1993) 445.
{\tt hep-ph/9210255}

\bibitem{CT}
K.G.~Chetyrkin, F.V.~Tkachov,
{\sl Nucl.~Phys.} {\bf B192} (1981) 159;\\
F.V.~Tkachov,
{\sl Phys.~Lett.} {\bf B100} (1981) 65.

\bibitem{IRA}
K.G.~Chetyrkin, F.V.~Tkachov,
{\sl Phys.~Lett.} {\bf B114} (1982) 340.

\bibitem{MM}
M.~Misiak, M.~M\"unz,
{\sl Phys.~Lett.} {\bf B344} (1995) 308.
{\tt hep-ph/9409454}

\bibitem{BG}
D.J.~Broadhurst, A.G.~Grozin,
{\sl Phys.~Rev.} {\bf D52} (1995) 4082.
{\tt hep-ph/9410240}

\bibitem{BK}
D.J.~Broadhurst, A.L.~Kataev,
{\sl Phys.~Lett.} {\bf B315} (1993) 179.
{\tt hep-ph/9308274}


\newpage

\begin{center}
{\bf Table~1}: Feynman-gauge contributions to $3\beta_3$
\end{center}
$$\begin{array}{|l|rr|}\hline
K_1K_1&24&\\\hline
K_1&-32&-48\zeta_3\\
K_1^{\prime\prime}&-4&\\
K_1^\prime&0&\\\hline
K_2&48&\\
K_2^{\prime\prime}&-36&+48\zeta_3\\
K_2^\prime&0&\\\hline
\mbox{total}&0&\\\hline
\end{array}$$

\begin{center}
{\bf Table~2}: Feynman-gauge contributions to $3\beta_4$
\end{center}
$$\begin{array}{|l|rrr|}\hline
K_1K_1K_1&48&&\\\hline
K_1K_1&-128&-192\zeta_3&\\
K_1K_1^\prime&0&&\\\hline
K_1K_2&192&&\\
K_1K_2^\prime&0&&\\\hline
K_{3a}&288&-2112\zeta_3&+1920\zeta_5\\
K_{3a}^{\prime\prime}&-494&-1600\zeta_3&+2400\zeta_5\\
K_{3a}^\prime&0&&\\\hline
K_{3b}&128&+1952\zeta_3&-2240\zeta_5\\
K_{3b}^{\prime\prime}&332&+2208\zeta_3&-2880\zeta_5\\
K_{3b}^\prime&128&+1536\zeta_3&-1920\zeta_5\\\hline
K_{3c}&-192&+672\zeta_3&-480\zeta_5\\
K_{3c}^{\prime\prime}&1264&+3040\zeta_3&-4800\zeta_5\\
K_{3c}^\prime&0&&\\\hline
K_{3d}&-160&-768\zeta_3&+960\zeta_5\\
K_{3d}^{\prime\prime}&-1222&-3136\zeta_3&+4800\zeta_5\\
K_{3d}^\prime&160&+1088\zeta_3&-1280\zeta_5\\\hline
K_1&520&+360\zeta_3&-480\zeta_5\\
K_1^{\prime\prime}&1656&+4632\zeta_3&-6880\zeta_5\\
K_1^\prime&-856&-1776\zeta_3&+2880\zeta_5\\\hline
K_2&-544&-1440\zeta_3&+1920\zeta_5\\
K_2^{\prime\prime}&-1120&-4464\zeta_3&+6080\zeta_5\\
K_2^\prime&-160&&\\\hline
\mbox{total}&-160&&\\\hline
\end{array}$$

\newpage

\begin{center}
{\bf Table~3}:
Dyson-Schwinger-Johnson anatomy of $\gamma_4$ by Gauss code
\end{center}
$$\begin{array}{|rcl|rrr|}
\hline
1&;&1&0&&\\
\hline
1,2&;&1,2&-\,{1459/8}&-\,20\,\zeta_5&+\,182\,\zeta_3\\
\hline
1,2,3&;&1,2,3&-\,{93/2}&+\,120\,\zeta_5&-\,114\,\zeta_3\\
1,2,3&;&1,3,2&183&-\,360\,\zeta_5&+\,96\,\zeta_3\\
1,2&;&3,1,2,3&29&-\,320\,\zeta_5&+\,314\,\zeta_3\\
1,2&;&3,2,1,3&598&-\,2440\,\zeta_5&+\,1642\,\zeta_3\\
\hline
1,2,3,1&;&4,2,3,4&-\,6&+\,20\,\zeta_5&-\,20\,\zeta_3\\
1,2,3,1&;&4,3,2,4&-\,70&+\,240\,\zeta_5&-\,144\,\zeta_3\\
1,2,3,4&;&1,2,3,4&-\,10&+\,20\,\zeta_5&-\,83\,\zeta_3\\
1,2,3,4&;&1,2,4,3&24&-\,120\,\zeta_5&+\,66\,\zeta_3\\
1,2,3,4&;&1,3,2,4&-\,1&-\,140\,\zeta_5&+\,119\,\zeta_3\\
1,2,3,4&;&1,3,4,2&5&&+\,9\,\zeta_3\\
1,2,3,4&;&1,4,2,3&-\,69&+\,120\,\zeta_5&-\,39\,\zeta_3\\
1,2,3,4&;&1,4,3,2&-\,26&+\,160\,\zeta_5&-\,112\,\zeta_3\\
1,2,3,4&;&2,1,4,3&-\,{15/2}&+\,260\,\zeta_5&-\,215\,\zeta_3\\
1,2,3,4&;&2,4,1,3&16&+\,160\,\zeta_5&-\,145\,\zeta_3\\
1,2,3&;&1,4,2,3,4&-\,27&+\,200\,\zeta_5&-\,131\,\zeta_3\\
1,2,3&;&1,4,3,2,4&53&-\,400\,\zeta_5&+\,295\,\zeta_3\\
1,2,3&;&2,4,1,3,4&56&-\,100\,\zeta_5&+\,43\,\zeta_3\\
1,2,3&;&2,4,3,1,4&-\,68&+\,40\,\zeta_5&+\,17\,\zeta_3\\
1,2,3&;&4,1,2,3,4&16&&+\,42\,\zeta_3\\
1,2,3&;&4,1,2,4,3&-\,25&+\,40\,\zeta_5&-\,7\,\zeta_3\\
1,2,3&;&4,1,3,2,4&-\,85&+\,360\,\zeta_5&-\,234\,\zeta_3\\
1,2,3&;&4,1,3,4,2&116&-\,200\,\zeta_5&+\,71\,\zeta_3\\
1,2,3&;&4,2,1,3,4&-\,19&-\,100\,\zeta_5&+\,106\,\zeta_3\\
1,2,3&;&4,2,1,4,3&79&-\,560\,\zeta_5&+\,407\,\zeta_3\\
1,2,3&;&4,2,3,1,4&37&-\,40\,\zeta_5&-\,5\,\zeta_3\\
1,2,3&;&4,3,1,2,4&125&-\,820\,\zeta_5&+\,595\,\zeta_3\\
1,2,3&;&4,3,1,4,2&-\,294&+\,1200\,\zeta_5&-\,795\,\zeta_3\\
1,2,3&;&4,3,2,1,4&-\,154&+\,840\,\zeta_5&-\,600\,\zeta_3\\
1,2&;&3,1,4,2,3,4&26&+\,400\,\zeta_5&-\,373\,\zeta_3\\
1,2&;&3,1,4,3,2,4&-\,130&+\,720\,\zeta_5&-\,507\,\zeta_3\\
1,2&;&3,2,4,1,3,4&-\,82&+\,960\,\zeta_5&-\,759\,\zeta_3\\
1,2&;&3,2,4,3,1,4&-\,58&+\,80\,\zeta_5&-\,17\,\zeta_3\\
1,2&;&3,4,1,2,3,4&5&-\,80\,\zeta_5&+\,101\,\zeta_3\\
1,2&;&3,4,1,2,4,3&-\,10&+\,80\,\zeta_5&-\,68\,\zeta_3\\
1,2&;&3,4,1,3,2,4&24&&-\,9\,\zeta_3\\
1,2&;&3,4,2,1,3,4&104&-\,400\,\zeta_5&+\,271\,\zeta_3\\
1,2&;&3,4,2,1,4,3&-\,122&+\,400\,\zeta_5&-\,244\,\zeta_3\\
1,2&;&3,4,2,3,1,4&-\,132&+\,320\,\zeta_5&-\,155\,\zeta_3\\
\hline
&&\mbox{total}&-\,{1027/8}&+\,640\,\zeta_5&-\,400\,\zeta_3\\
\hline
\end{array}$$

\newpage

\begin{center}
{\bf Table~4}:
Dyson-Schwinger-Johnson anatomy of $\gamma_{m,4}+\gamma_4$ by Gauss code
\end{center}
$$\begin{array}{|rcl|rrr|}
\hline
1&;&1&-\,602&+\,1280\,\zeta_5&-\,992\,\zeta_3\\
\hline
1,2&;&1,2&212&-\,320\,\zeta_5&+\,632\,\zeta_3\\
\hline
1,2,3&;&1,2,3&144&+\,320\,\zeta_5&-\,80\,\zeta_3\\
1,2,3&;&1,3,2&-\,848&+\,1600\,\zeta_5&-\,928\,\zeta_3\\
1,2&;&3,1,2,3&-\,1120&-\,2560\,\zeta_5&+\,2272\,\zeta_3\\
1,2&;&3,2,1,3&-\,272&+\,3520\,\zeta_5&-\,2656\,\zeta_3\\
\hline
1,2,3,1&;&4,2,3,4&-\,16&+\,960\,\zeta_5&-\,696\,\zeta_3\\
1,2,3,1&;&4,3,2,4&200&&-\,96\,\zeta_3\\
1,2,3,4&;&1,2,3,4&1056&-\,800\,\zeta_5&-\,112\,\zeta_3\\
1,2,3,4&;&1,2,4,3&-\,1664&+\,2560\,\zeta_5&-\,640\,\zeta_3\\
1,2,3,4&;&1,3,2,4&-\,480&+\,480\,\zeta_5&+\,96\,\zeta_3\\
1,2,3,4&;&1,3,4,2&-\,672&&+\,672\,\zeta_3\\
1,2,3,4&;&1,4,2,3&-\,992&+\,160\,\zeta_5&+\,512\,\zeta_3\\
1,2,3,4&;&1,4,3,2&2336&-\,960\,\zeta_5&-\,960\,\zeta_3\\
1,2,3,4&;&2,1,4,3&1008&-\,1440\,\zeta_5&+\,384\,\zeta_3\\
1,2,3,4&;&2,4,1,3&-\,224&&+\,384\,\zeta_3\\
1,2,3&;&1,4,2,3,4&384&-\,3520\,\zeta_5&+\,2560\,\zeta_3\\
1,2,3&;&1,4,3,2,4&-\,256&+\,2880\,\zeta_5&-\,2208\,\zeta_3\\
1,2,3&;&2,4,1,3,4&-\,128&-\,960\,\zeta_5&+\,864\,\zeta_3\\
1,2,3&;&2,4,3,1,4&256&-\,320\,\zeta_5&-\,160\,\zeta_3\\
1,2,3&;&4,1,2,3,4&-\,1664&+\,3200\,\zeta_5&-\,1664\,\zeta_3\\
1,2,3&;&4,1,2,4,3&800&-\,2240\,\zeta_5&+\,992\,\zeta_3\\
1,2,3&;&4,1,3,2,4&928&-\,2880\,\zeta_5&+\,1536\,\zeta_3\\
1,2,3&;&4,1,3,4,2&-\,288&+\,1600\,\zeta_5&-\,1216\,\zeta_3\\
1,2,3&;&4,2,1,3,4&1280&-\,1120\,\zeta_5&-\,224\,\zeta_3\\
1,2,3&;&4,2,1,4,3&-\,976&&+\,768\,\zeta_3\\
1,2,3&;&4,2,3,1,4&288&+\,1600\,\zeta_5&-\,1696\,\zeta_3\\
1,2,3&;&4,3,1,2,4&128&+\,800\,\zeta_5&-\,608\,\zeta_3\\
1,2,3&;&4,3,1,4,2&352&&-\,480\,\zeta_3\\
1,2,3&;&4,3,2,1,4&-\,800&-\,960\,\zeta_5&+\,1344\,\zeta_3\\
1,2&;&3,1,4,2,3,4&-\,640&+\,960\,\zeta_5&-\,192\,\zeta_3\\
1,2&;&3,1,4,3,2,4&800&&-\,672\,\zeta_3\\
1,2&;&3,2,4,1,3,4&640&-\,1280\,\zeta_5&+\,512\,\zeta_3\\
1,2&;&3,2,4,3,1,4&-\,192&&+\,288\,\zeta_3\\
1,2&;&3,4,1,2,3,4&-\,352&-\,2240\,\zeta_5&+\,2336\,\zeta_3\\
1,2&;&3,4,1,2,4,3&1072&+\,1920\,\zeta_5&-\,2496\,\zeta_3\\
1,2&;&3,4,1,3,2,4&-\,224&+\,1600\,\zeta_5&-\,928\,\zeta_3\\
1,2&;&3,4,2,1,3,4&-\,128&+\,1600\,\zeta_5&-\,1216\,\zeta_3\\
1,2&;&3,4,2,1,4,3&-\,144&-\,2560\,\zeta_5&+\,2464\,\zeta_3\\
1,2&;&3,4,2,3,1,4&512&-\,2240\,\zeta_5&+\,1568\,\zeta_3\\
\hline
&&\mbox{total}&-\,286&+\,640\,\zeta_5&-\,736\,\zeta_3\\
\hline
\end{array}$$

\end{thebibliography}
\end{document}